\documentclass[12pt]{article}
\usepackage{amsfonts}
\usepackage[dvips]{epsfig}
\usepackage{graphicx}
\begin{document}

\title{Topology, Quasiperiodic functions and the Transport
phenomena.}

\author{A.Ya.Maltsev$^{1}$, S.P.Novikov$^{1,2}$}

\date{
\centerline{$^{(1)}$ L.D.Landau Institute for Theoretical Physics,}
\centerline{119334 ul. Kosygina 2, Moscow, }
\centerline{ maltsev@itp.ac.ru \,\, ,
\,\,\, novikov@itp.ac.ru}
\centerline{$^{(2)}$ IPST, University of Maryland,}
\centerline{College Park MD 20742-2431,USA}
\centerline{novikov@ipst.umd.edu}}

\maketitle

\begin{abstract}
 In this article we give the basic concept of the 
"Topological Numbers" in theory of quasiperiodic functions.
The main attention is paid to apperance of such values
in transport phenomena
including Galvanomagnetic phenomena in normal metals (Chapter 1)
and the modulations of 2D electron gas (Chapter 2). We give just 
the main introduction to both of these areas and explain in a 
simple way the appearance of the "integral characteristics"
in both of these problems. The paper can not be considered
as the detailed survey article in the area but explains the
main basic features of the corresponding phenomena.
\end{abstract}

\centerline{{\bf {\Large Introduction.}}}

\begin{center}
{\bf {\Large Galvanomagnetic phenomena in normal metals,
Transport in 2D electron gas and Topology of Quasiperiodic functions.}}
\end{center}

\vspace{0.5cm}

 We are going to consider the transport phenomena connected with the
geometry of quasiclassical electron trajectories in the magnetic 
field ${\bf B}$. 

\begin{figure}
\begin{center}
\epsfig{file=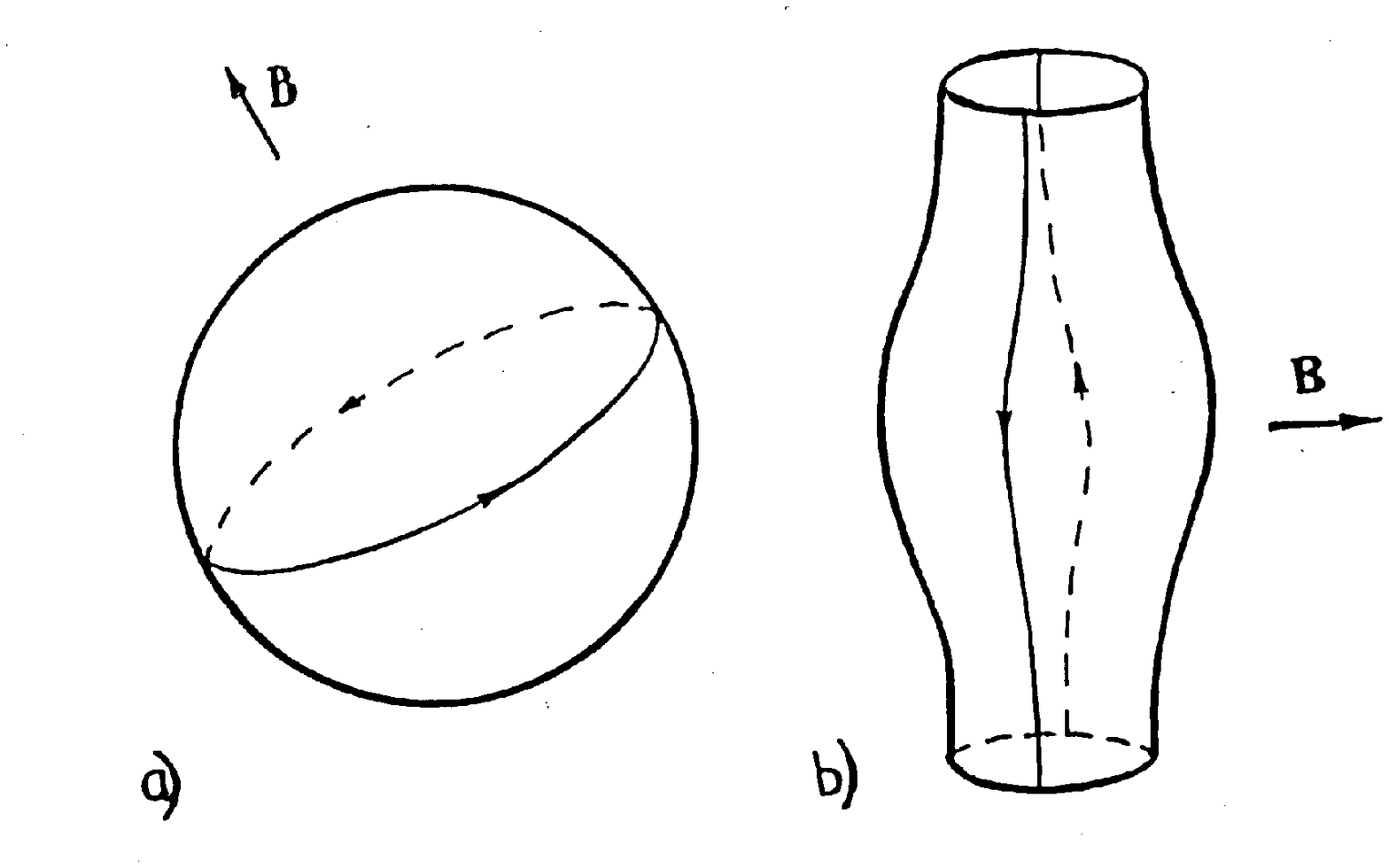,width=14.0cm,height=8cm}
\end{center}
\caption{The "simple" Fermi surface having the form of the sphere
in the Brillouen zone and the periodic "warped cylinder" extending
through the infinite number of Brillouen zones. The quasiclassical 
electron orbits in ${\bf p}$-space are also shown for a given 
direction of ${\bf B}$.
}
\label{drezden1}
\end{figure}

 Let us start with the most fundamental case where
this kind of phenomena appears in the conductivity of normal metals
having complicated Fermi surfaces in the presence of the rather 
strong magnetic field. This classical part of of the solid state
physics was started by Kharkov school of I.M. Lifshitz 
(I.M. Lifshitz, M.Ya. Azbel, M.I. Kaganov, V.G. Peschansky)
in 1950's and has become the essential part of conductivity theory
in normal metals. Let us give here some small excurse in this area.
We will start with the classical work of I.M. Lifshitz, M.Ya. Azbel
and M.I. Kaganov (\cite{lifazkag}) where the importance of
topology of the Fermi surface for the conductivity was established.
Namely, there was shown the difference between the "simple" Fermi
surface (topological "sphere") (Fig. \ref{drezden1},a) and more 
complicated surfaces where the non-closed quasiclassical electron 
trajectories can arise. In particular, the detailed consideration 
for the "simple" Fermi surface and the surfaces like "warped cylinder" 
(Fig. \ref{drezden1},b) for the different directions of ${\bf B}$ 
was made.

 Both the pictures on Fig. \ref{drezden1} represent the forms of 
the Fermi surfaces in ${\bf p}$-space and we should remember that 
only one Brillouen zone should be taken in the account to get the
right phase space volume for the electron states. The values of 
${\bf p}$ different by any reciprocal lattice vector
$n_{1} {\bf a}_{1} + n_{2} {\bf a}_{2} + n_{3} {\bf a}_{3}$
(where $n_{i}$ are integers) are then physically equivalent to each
other and represent the same electron state. The Brillouen zone
can then be considered as the parallelogram in the 
${\bf p}$-space with the identified opposite sides on the boundary.

\begin{figure}
\begin{center}
\epsfig{file=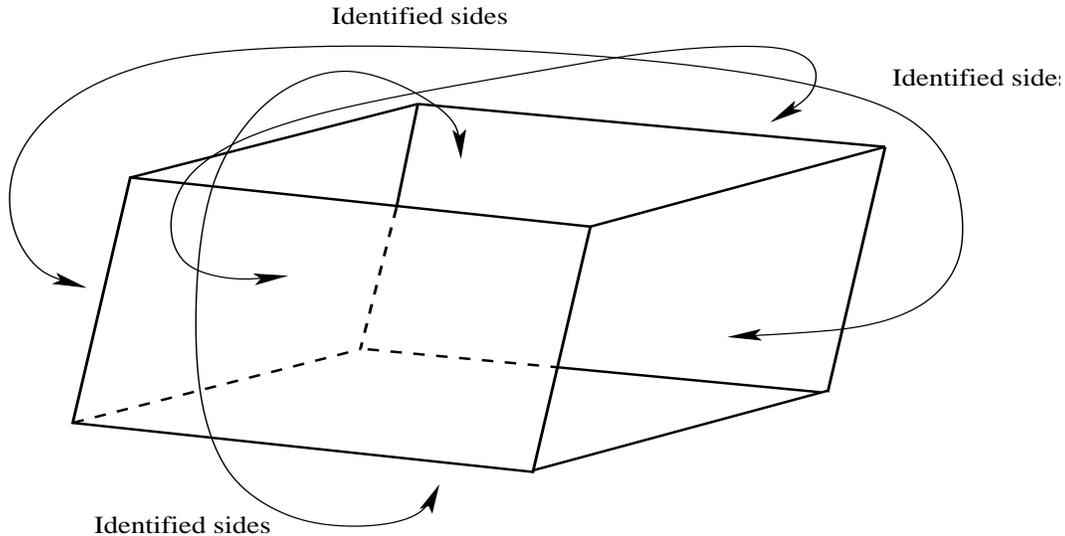,width=14.0cm,height=7cm}
\end{center}
\caption{The Brillouen zone in the quasimomentum ({\bf p}) space
with the identifies sides on the boundary.
}
\label{drezden2}
\end{figure}

 Also the Fermi surfaces $S_{F}$ will then be periodic in 
${\bf p}$-space with periods ${\bf a}_{1}$, ${\bf a}_{2}$, 
${\bf a}_{3}$.

 {\bf Remark.} From topological point of view we can consider the
Brillouen zone as the compact 3-dimensional torus ${\mathbb T}^{3}$.
The corresponding Fermi-surfaces will then be also compact surfaces
of finite size embedded in ${\mathbb T}^{3}$.

 The presence of the homogeneous magnetic field ${\bf B}$ generates 
the evolution of electron states in the ${\bf p}$-space which can be
described by the dynamical system

\begin{equation}
\label{dynsyst}
{\dot {\bf p}} = {e \over c} \left[ {\bf v}_{gr}({\bf p}) \times
{\bf B} \right] = {e \over c} \left[ \nabla \epsilon({\bf p}) \times
{\bf B} \right]
\end{equation}
where $\epsilon({\bf p})$ is the dependence of energy on the 
quasimomentum (dispersion relation) and 
${\bf v}_{gr}({\bf p}) = \nabla \epsilon({\bf p})$ is the group
velocity at the state ${\bf p}$. Both the functions $\epsilon({\bf p})$
and ${\bf v}_{gr}({\bf p})$ are also the periodic functions in 
${\bf p}$-space and can be considered as the one-valued functions
in ${\mathbb T}^{3}$.

 The system (\ref{dynsyst}) has two conservative integrals which are
the electron energy and the component of ${\bf p}$ along the magnetic
field. The electron trajectories can then be represented as the 
intersections
of the constant energy surfaces $\epsilon({\bf p}) = const$ with the
planes orthogonal to ${\bf B}$ and only the Fermi level 
$\epsilon({\bf p}) = \epsilon_{F}$ is actually important for the 
conductivity. Easy to see then that global geometry of the "essential"
electron trajectories will depend strongly on the form of Fermi surface
in ${\bf p}$-space.

 Coming back to the Fig. \ref{drezden1} we can see that the form of 
electron trajectories can be quite different for the Fermi surfaces
shown at Fig. \ref{drezden1},a and Fig. \ref{drezden1},b. Such,
for the Fermi surface shown at Fig. \ref{drezden1},b we can have
periodic non-closed electron trajectories (if ${\bf B}$ is orthogonal
to vertical axis) while for the surface on Fig.  \ref{drezden1},a
all the trajectories are just closed curves lying in one Brillouen 
zone for all directions of ${\bf B}$.

\begin{figure}
\begin{center}
\epsfig{file=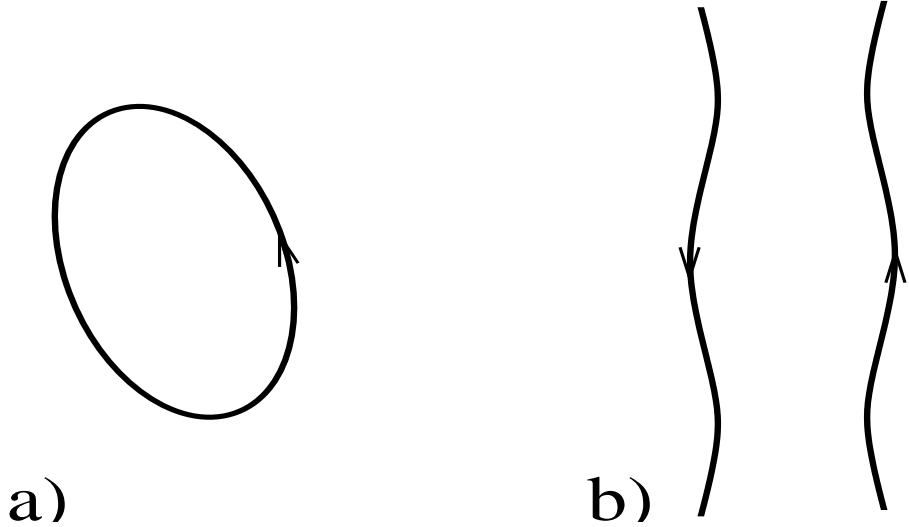,width=12.0cm,height=7cm}
\end{center}  
\caption{Electron trajectories in ${\bf p}$-space given by the
intersections of planes orthogonal to ${\bf B}$ for the Fermi surfaces
shown at Fig. \ref{drezden1}a and Fig. \ref{drezden1}b for ${\bf B}$
orthogonal to vertical axis.
}
\label{drezden3}
\end{figure}

 Let us say now that this global geometry plays the main role in the 
electron motion in the coordinate space also (despite the factorization
in ${\bf p}$-space). Thus the electron wave-packet motion in
${\bf x}$-space (${\bf x} = (x, y, z)$) can be found from the additional 
system
 
$${\dot {\bf x}} = {\bf v}_{gr}({\bf p}(t)) = 
\nabla \epsilon({\bf p}(t))$$
for any trajectory in ${\bf p}$-space after the integration of system
(\ref{dynsyst}). The structure of system (\ref{dynsyst}) permits to 
claim for example that the $xy$-projection of "electron motion" in
${\bf x}$-space has the same form as the trajectory in ${\bf p}$-space
just rotated by $\pi/2$. We can see then that the electron drift in
${\bf x}$-space in magnetic field is also very different for the
trajectories shown at Fig. \ref{drezden3}a and \ref{drezden3}b 
due to the action of the crystal lattice.
 
 The effect of this "geometrical drift" can be measured experimentally
in the rather pure metallic monocrystals if the mean free electron
motion time is big enough (such that electron packet "feels" the 
geometric features of trajectory between the two scattering acts).
The geometric picture requires then that the time between the two 
scatterings is much longer than the "passing time" through one 
Brillouen  zone for the periodic trajectory and much longer
than the "inverse cyclotron frequency" for closed 
trajectories.\footnote{This criterium
can be actually more complicated for trajectories of more complicated
form.} For the approximation of effective mass $m^{*}$ in crystal this
condition can be roughly expressed as $\omega_{B} \tau \gg 1$, where
$\omega_{B} = eB/m^{*}c$ is the formal cyclotron frequency and 
$\tau$ is the mean free electron motion time. Let us note that this
requirement is satisfied better for the big values of $B$ and we will 
consider the formal limit $B \rightarrow \infty$ in our paper. We
will call this situation "geometric strong magnetic field limit"
and consider the asymptotic of conductivity tensor for this 
case.\footnote{Formally also another condition 
$\hbar \omega_{B} \ll \epsilon_{F}$ should also be imposed on the
magnetic field $B$. However, this condition is always satisfied for 
the real metals and all experimentally available magnetic fields
(the upper limit is $B \sim 10^{3}-10^{4} Tl$). So we will not pay
special attention to this second restriction and assume that the 
limit $B \rightarrow \infty$ is considered in the "experimental
sence" where the second condition is satisfied.}

 Let us give here the asymptotic form of conductivity tensor
obtained in \cite{lifazkag} for the case of trajectories shown
at Fig. \ref{drezden3}a and Fig. \ref{drezden3}b. Let us take 
the $z$-axis in the
${\bf x}$-space along the magnetic field ${\bf B}$. The axes $x$
and $y$ can be chosen arbitrarily for the case of
Fig. \ref{drezden3}a and we take the $y$-axis along the mean 
electron drift direction in ${\bf x}$-space for the case 
Fig. \ref{drezden3}b. (Easy to see that the $x$-axis will then be 
directed along the mean electron drift in ${\bf p}$-space in this 
situation). The asymptotic forms of the conductivity tensor can then 
be written as:

\vspace{0.5cm} 
    
 {\bf Case 1} (Closed trajectories, Fig. \ref{drezden3}a):
  
\begin{equation}
\label{sigmacl}
\sigma^{ik} \simeq {n e^{2} \tau \over m^{*}} \,
\left( \begin{array}{ccc}
(\omega_{B}\tau)^{-2} & (\omega_{B}\tau)^{-1} &
(\omega_{B}\tau)^{-1} \cr
(\omega_{B}\tau)^{-1} & (\omega_{B}\tau)^{-2} &
(\omega_{B}\tau)^{-1} \cr
(\omega_{B}\tau)^{-1} & (\omega_{B}\tau)^{-1} & *
\end{array} \right) \,\,\, , \,\,\, \omega_{B}\tau \gg 1
\end{equation}

\vspace{0.5cm}
 
 {\bf Case 2} (open periodic trajectories,  Fig. \ref{drezden3}b):

\begin{equation}
\label{sigmaop}
\sigma^{ik} \simeq {n e^{2} \tau \over m^{*}} \,
\left( \begin{array}{ccc}
(\omega_{B}\tau)^{-2} & (\omega_{B}\tau)^{-1} &
(\omega_{B}\tau)^{-1} \cr
(\omega_{B}\tau)^{-1} & * & * \cr
(\omega_{B}\tau)^{-1} & * & *
\end{array} \right) \,\,\, , \,\,\, \omega_{B}\tau \gg 1
\end{equation}
where $*$ mean some dimensionless constants of order of 1.
 
\vspace{0.5cm}

 We can see that the conductivity reveals the strong anisotropy
in the plane orthogonal to ${\bf B}$ in the second case and the 
mean direction of the electron trajectory in ${\bf p}$-space
(not in ${\bf x}$) can be measured experimentally as the zero
eigen-direction of $\sigma^{ik}$ for $B \rightarrow \infty$.

\begin{figure}
\begin{center}
\epsfig{file=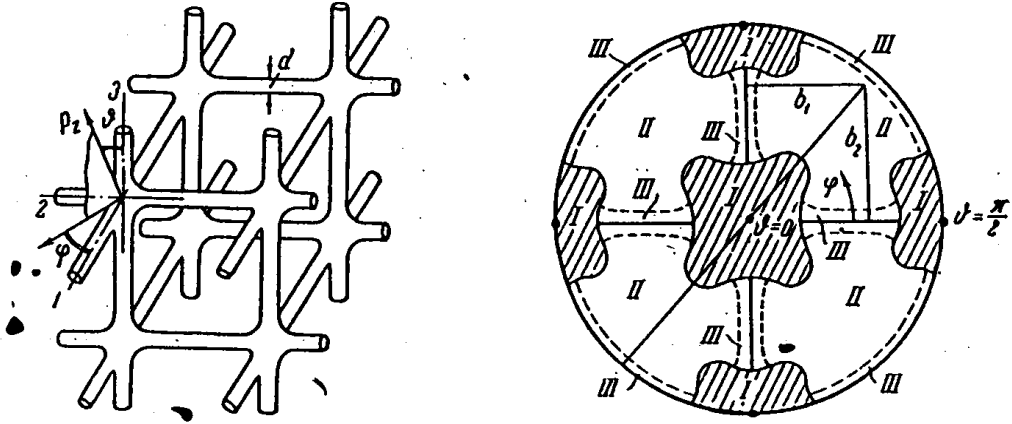,width=14.0cm,height=6cm}
\end{center}
\caption{The picture from \cite{lifpes1} representing the 
"thin spatial net" and the corresponding directions of ${\bf B}$
on the unit sphere where the non-closed electron trajectories 
exist.
}
\label{drezden4}
\end{figure}

 More general types of open electron trajectories were considered
in \cite{lifpes1,lifpes2}. For example, the open trajectories which
are not periodic were found in  \cite{lifpes1} for the 
"thin spatial net" (Fig. \ref{drezden4}, a).

 The open trajectories exist here only for the directions of 
${\bf B}$ close to main crystallographic axes $(1, 0, 0)$,
$(0, 1, 0)$ and $(0, 0, 1)$ (Fig. \ref{drezden4}, b). It was
shown in \cite{lifpes1} that the open trajectories lie 
in this case in the straight strip of the finite width in the plane 
orthogonal to ${\bf B}$ and pass through them. 
The mean direction of open trajectories are given here by the
intersections of plane orthogonal to ${\bf B}$ with the main
crystallographic planes $(xy)$, $(yz)$ and $(xz)$.

 The form of conductivity tensor for this kind of trajectories 
used in \cite{lifpes1} coincides with (\ref{sigmaop}).

 Some analytical dispersion relations were also considered in
\cite{lifpes2}.\footnote{Actually this work contains some 
conceptual mistakes but gives also some correct features concerning
the existence of open trajectories for these dispersion relations.}
Let us mention here also the works 
\cite{aleksgaid1,aleksgaid2,aglifpes,aleksgaid3,gaid1,aleksgaid4,
aleksgaid5,aleksgaid6} where different experimental (and theoretical)
investigations for some real metals were made. The detailed 
consideration of these results can be found also in the survey 
articles \cite{lifkag1, lifkag2} and the book \cite{etm}
(see also \cite{abr}).

\vspace{0.5cm}

 Let us say now about the topological approach to the problem of
general classification of all possible electron trajectories 
regardless the concrete features of the dispersion relation
$\epsilon({\bf p})$ started by S.P. Novikov (\cite{novikov1})
(see also \cite{novikov2,novikov3,novikov4}). Let
us formulate here the Novikov problem:

\vspace{0.5cm}

 {\bf Novikov problem.} {\it Let any smooth 3-periodic function 
$\epsilon({\bf p})$ be given in the 3-dimensional space 
${\mathbb R}^{3}$ (with arbitrary lattice of periods). Fix any
non-degenerate energy level $\epsilon({\bf p}) = const$
(i.e. $\nabla \epsilon({\bf p}) \neq 0$ on this level) and
consider the intersections of corresponding smooth 3-periodic
surface by any set of parallel planes in ${\mathbb R}^{3}$.
Describe the global geometry of all possible non-singular 
(open) trajectories which can arise in the intersections.}

\vspace{0.5cm}

 The words "the global geometry" mean here first of all the 
asymptotic behavior of the trajectory when $t \rightarrow \pm \infty$
in sence of dynamical systems. 
Let us formulate here also Novikov conjecture about the generic 
non-singular trajectories which was proved later by his pupils:

\vspace{0.5cm}

 {\bf Novikov conjecture.} {\it The generic non-singular open 
trajectories lie in the straight strips of finite width (in the
plane orthogonal to ${\bf B}$) and pass through them.}

\vspace{0.5cm}

 Let us emphasize also that Novikov conjecture is connected with
the generic open trajectories and can be not valid in the special
degenerate cases (S.P. Tsarev, I.A. Dynnikov) as we will see
later.

 The topological problem of S.P. Novikov was considered later in
his school (A.V. Zorich, I.A. Dynnikov, S.P. Tsarev) where the 
basic theorems about the non-closed trajectories were obtained.
Let us say here about the main breakthroughs in this problem
made in \cite{zorich1} (A.V. Zorich) and \cite{dynn3} 
(I.A. Dynnikov).

 We note first that even for the rather complicated periodic Fermi 
surface the electron trajectories will be quite simple if the 
direction of ${\bf B}$ is purely rational (with respect to
reciprocal lattice), i.e. if the plane $\Pi({\bf B})$ orthogonal to
${\bf B}$ contains two linearly independent reciprocal lattice
vectors. This property can also be formulated in the form that the 
magnetic fluxes through the faces of elementary cell in the 
${\bf x}$-space are proportional to each other with rational
coefficients. In this situation the picture arising in 
$\Pi({\bf B})$ is purely periodic and all open electron 
trajectories can be also just the periodic curves corresponding
precisely to the case (\ref{sigmaop}). However, the condition of
rationality is completely unstable with respect to any small 
rotations of ${\bf B}$ such that the rational directions give 
just a set of measure zero among all the directions of ${\bf B}$.

 The remarkable fact proved by A.V. Zorich is that the open 
trajectories reveal the "topologically regular" properties even after 
the small rotations purely rational direction. Namely, they lie in 
the straight strips of the finite width in accordance with Novikov 
conjecture (but are not periodic anymore) and pass through them. 
Let us formulate this in more precise form.

\vspace{0.5cm}

\begin{figure}
\begin{center}
\epsfig{file=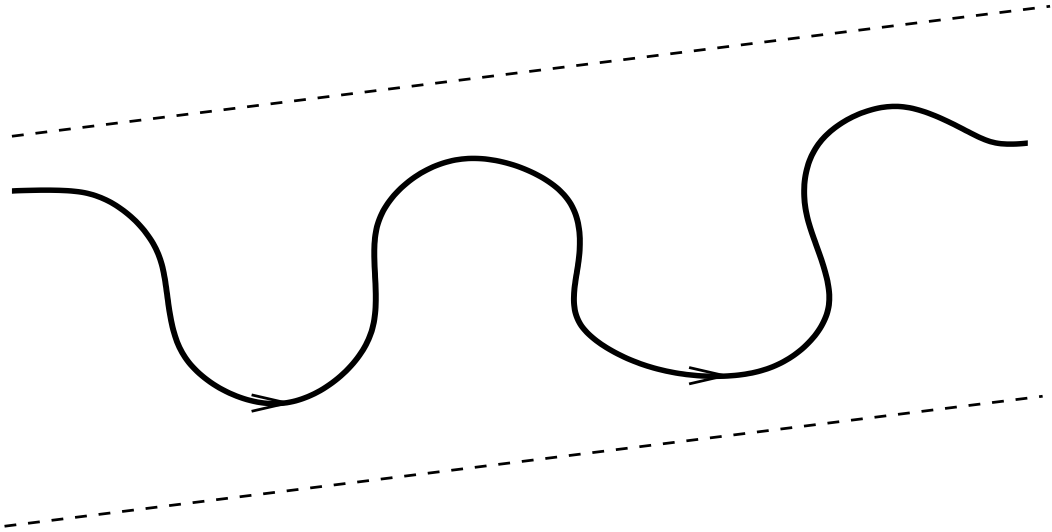,width=14.0cm,height=7cm}
\end{center}
\caption{General open trajectory lying in the straight strip
of finite width in the plane orthogonal to ${\bf B}$.
}
\label{drezden6}
\end{figure}

{\bf Theorem 1.} (A.V. Zorich, \cite{zorich1}) {\it
Consider arbitrary smooth Fermi surface and the rational direction
of magnetic field ${\bf B}_{0}$ such that no singular trajectory 
connects two different (not equivalent modulo the reciprocal lattice)
singular (stagnation) points of the system (\ref{dynsyst}). Then 
there exists small open region $\Omega$ on the unit sphere around
direction ${\bf B}_{0}$ such that all open trajectories 
(if they exist) lie in the straight strips of finite width in the
plane orthogonal to ${\bf B}$ if ${\bf B}/B \in \Omega$,
(Fig. \ref{drezden6}).}

\vspace{0.5cm}

 Let us mention also that the additional topological condition in
Theorem 1 has a generic form and generically does not impose 
anything on the direction ${\bf B}_{0}$.

 Theorem of Zorich claims actually that all the rational directions
of ${\bf B}$ can be extended to some "small open spots" on the
unit sphere (parameterizing directions of ${\bf B}$ where we can not 
have situation more complicated than represented at 
Fig. \ref{drezden6}. This set already has the finite measure on the 
unit sphere and moreover we can conclude that any stable open
trajectory can have only the form shown at Fig. \ref{drezden6}
since the rational directions are everywhere dense on the unit
sphere. Zorich theorem, however, does not permit to state that
this situation is the only possible one since the sizes of the
"spots" become smaller and smaller for big rational numbers and
we can not claim that they cover all the unit sphere in general 
situation.

\vspace{0.5cm}

 The next important result was obtained by I.A. Dynnikov 
(\cite{dynn3}) who proved that the trajectories shown at 
Fig. \ref{drezden6} can be the only stable trajectories with
respect to the small variation of the Fermi energy $\epsilon_{F}$
for a given dispersion relation $\epsilon({\bf p})$. Let us 
formulate the exact form Dynnikov theorem in Chapter 1 where
we will consider this picture in more details. We will just say 
here that the methods developed in \cite{dynn3} permitted to prove 
later that all the cases of open trajectories different from shown
at Fig. \ref{drezden6} can appear only "with probability zero"
(i.e. for the directions of ${\bf B}$ from the set of measure zero
on the unit sphere) for generic Fermi surfaces 
$S_{F}: \, \epsilon({\bf p}) = \epsilon_{F}$ (\cite{dynn4,dynn7})
which gave the final proof of Novikov conjecture for generic
open trajectories.

\vspace{0.5cm}

 The methods of proofs of Zorich and Dynnikov theorems gave the 
basis for the invention of the "Topological Quantum Numbers" 
introduced in \cite{novmal1} by present authors (see also the
survey articles \cite{novmal2,malnov3,malnov4}) for the conductivity
in normal metals. Let us say also that another important property,
called later the "Topological Resonance" played the crucial role
for physical phenomena in \cite{novmal1}. The main point of this 
property can be formulated as follows: all the trajectories having 
the form shown at Fig. \ref{drezden6} have the same mean direction
in all the planes orthogonal to ${\bf B}$ for the generic directions
of ${\bf B}$ (actually for any not purely rational direction of 
${\bf B}$) and give the same form (\ref{sigmaop}) of contribution 
to conductivity tensor in the same coordinate system. This important
fact makes experimentally observable the integer-valued topological
characteristics of the Fermi surface having the form of the
integral planes of reciprocal lattice and corresponding 
"stability zones" on the unit sphere. We are going to describe in 
details these quantities in the Chapter 1 of our paper. Our goal
is to give here the main features of the corresponding picture and we 
don't give all the details of the classification of all open 
trajectories for general Fermi surfaces. However, the picture 
we are going to describe serves as the "basic description" of
conductivity phenomena and all the other possibilities can be 
considered as the special additional features for the non-generic
directions of ${\bf B}$. Let us also say here that the final 
classification of open trajectories for generic Fermi surfaces
was finished in general by I.A. Dynnikov in \cite{dynn7} which solves
in main the Novikov problem. The physical phenomena connected with
different types of open trajectories can be found in details in
the survey articles \cite{malnov3,malnov4}.

\vspace{0.5cm}

 Let us say now some words about the general Novikov problem connected
with the quasiperiodic functions on the plane with $N$ quasiperiods.
According to the standard definition the quasiperiodic function in
${\mathbb R}^{m}$ with $N$ quasiperiods ($N \geq m$) is a restriction
of a periodic function in ${\mathbb R}^{N}$ (with $N$ periods) to any
plane ${\mathbb R}^{m} \subset {\mathbb R}^{N}$ of dimension $m$
linearly embedded in ${\mathbb R}^{N}$. In our situation we will
always have $m = 2$ and the quasiperiodic functions on the plane will
be the restrictions of the periodic functions in ${\mathbb R}^{N}$ to
some 2D plane.

\vspace{0.5cm}

 {\bf General Novikov problem}. {\it Describe the global geometry
of open level curves of quasiperiodic function $f({\bf r})$ on
the plane with $N$ quasiperiods.}

\vspace{0.5cm}

 Easy to see that the general Novikov problem gives the Novikov 
problem for the electron trajectories if we put $N = 3$. Indeed, 
all the trajectories in the planes orthogonal to ${\bf B}$ can be
considered as the level curves of quasiperiodic functions
$\epsilon({\bf p})|_{\Pi({\bf B})}$ with 3 quasiperiods. 
According to the said above we can say that the general Novikov 
problem is solved in main for $N = 3$. However, the case $N > 3$
becomes very complicated from topological point of view and no
general classification in this case exists at the moment. Let us say 
that the only topological result existing now for general Novikov 
problem is the analog of Zorich theorem (Theorem 1) for the case
$N = 4$ (S.P. Novikov, \cite{novikov5}) and the general situation
is still under investigation by now.

 In Chapter 2 of our paper we consider another application of
general Novikov problem connected with the "superlattice potentials"
for the two-dimensional electron gas in the presence of orthogonal 
magnetic field. This kind of potentials is connected with modern
techniques of "handmade" modulations of 2D electron gas such as the
holographic illumination, "gate modulation", piezoelectric effect
etc ... . All such modulations are usually periodic in the plane
and in many situations the level curves play the important role for
the transport phenomena in such systems. The most important thing for 
us will be the conductivity phenomena in these 2D structures in the
presence of orthogonal magnetic field ${\bf B}$. According to the 
quasiclassical approach the cyclotron electron orbits drift along the
level curves of modulation potential in the magnetic field which 
gives the "drift contribution" to conductivity in the plane. 
Among the papers devoted to this approach we would like to mention 
here the paper \cite{beenakker} (C. Beenakker) where this approach was 
introduced for 
the explanation of "commensurability oscillations" of conductivity 
in potential modulated just in one direction and
\cite{GrLonDav} (D.E. Grant, A.R. Long, J.H. Davies) 
where the same approach was used for explanation
of suppression of these oscillations by the second orthogonal 
modulation in the periodic case. Let us add that all these phenomena
correspond to the long free electron motion time which will now play
the role of the "geometric limit" (not $B \rightarrow \infty$) in
this second situation.

 We are going to show that the general Novikov problem can also
arise naturally in these structures if we consider the independent 
superposition of different periodic modulations. It can be proved
that in this case we always obtain the quasiperiodic functions where
the number of quasiperiods depends on the complexity of total 
modulation. The results in Novikov problem can then help to predict
the form of the "drift conductivity" in the limit of long free electron
motion time. In Chapter 2 we give the main features of the situation 
of superposition of several "1D modulations" where the potentials with 
small number of quasiperiods can arise. The detailed consideration of
this situation can be found in \cite{maltsev2}. However, the 
Novikov problem arise also in much more general case of arbitrary
superpositions of more complicated (but periodic) structures.

 At last we would like to say that the quasiperiodic functions
with big number of quasiperiods can be a model for the random 
potentials on the plane. The corresponding Novikov problem arise
in the percolation theory for such potentials. We will also say some 
words about this situation at the end of Chapter 2.

\section{The classification of Fermi surfaces and the "Topological
Quantum Numbers".}

 Let us start with the definitions of genus and Topological Rank
of the Fermi surface.

\vspace{0.5cm}

 {\bf Definition 1.}

\begin{figure}
\begin{center}
\epsfig{file=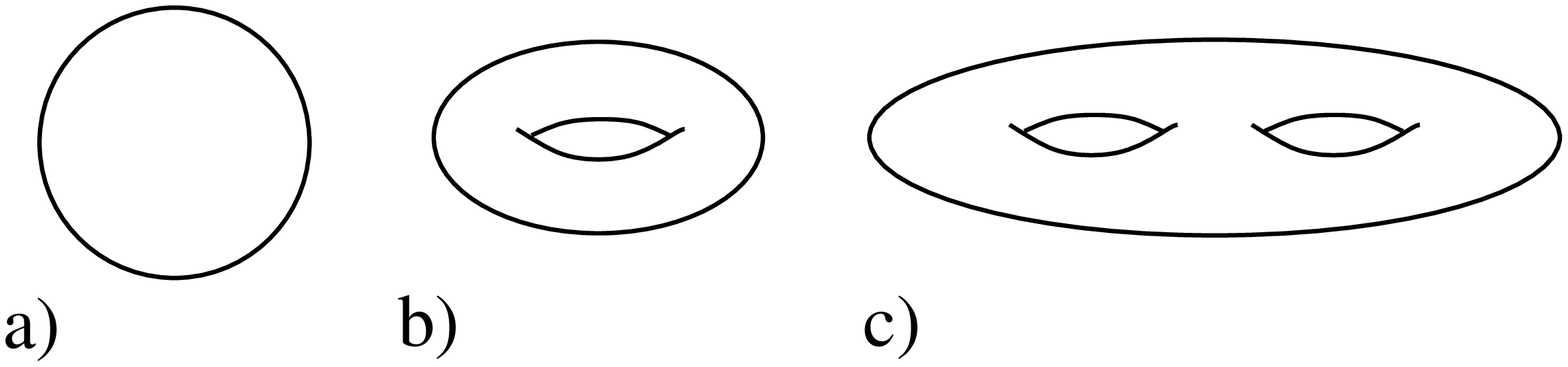,width=14.0cm,height=3.5cm}
\end{center}
\caption{The abstract surfaces with genuses $0$, $1$ and $2$
respectively.
}
\label{drezden7}
\end{figure}

{\it  Let us consider the phase space
${\mathbb T}^{3} = {\mathbb R}^{3}/L$ introduce above. After
the identification
every component of the Fermi surface becomes the smooth
orientable 2-dimensional surface embedded in ${\mathbb T}^{3}$.
We can then
introduce the standard genus of every component of the Fermi
surface $g = 0,1,2,...$ according to standard topological
classification depending on if this component is topological
sphere, torus, sphere with two holes, etc ... (Fig. \ref{drezden7})}

\vspace{0.5cm}

\vspace{0.5cm}

 {\bf Definition 2.}

{\it Let us introduce the Topological Rank $r$ as the 
characteristic of the embedding of the Fermi surface in 
${\mathbb T}^{3}$. It's much more
convenient in this case to come back to the total ${\bf p}$-space
and consider the connected components of the three-periodic
surface in ${\mathbb R}^{3}$. 

1) The Fermi surface has Rank $0$ if every its connected component
can be bounded by a sphere of finite radius.

2) The Fermi surface has Rank $1$ if every its connected component
can be bounded by the periodic cylinder of finite radius and there
are components which can not be bounded by the sphere.

3) The Fermi surface has Rank $2$ if every its connected component
can be bounded by two parallel (integral) planes in
${\mathbb R}^{3}$ and
there are components which can not be bounded by cylinder.

4) The Fermi surface has Rank $3$ if it contains components which
can not be bounded by two parallel planes in
${\mathbb R}^{3}$. }

\vspace{0.5cm}

\begin{figure}
\begin{center}
\epsfig{file=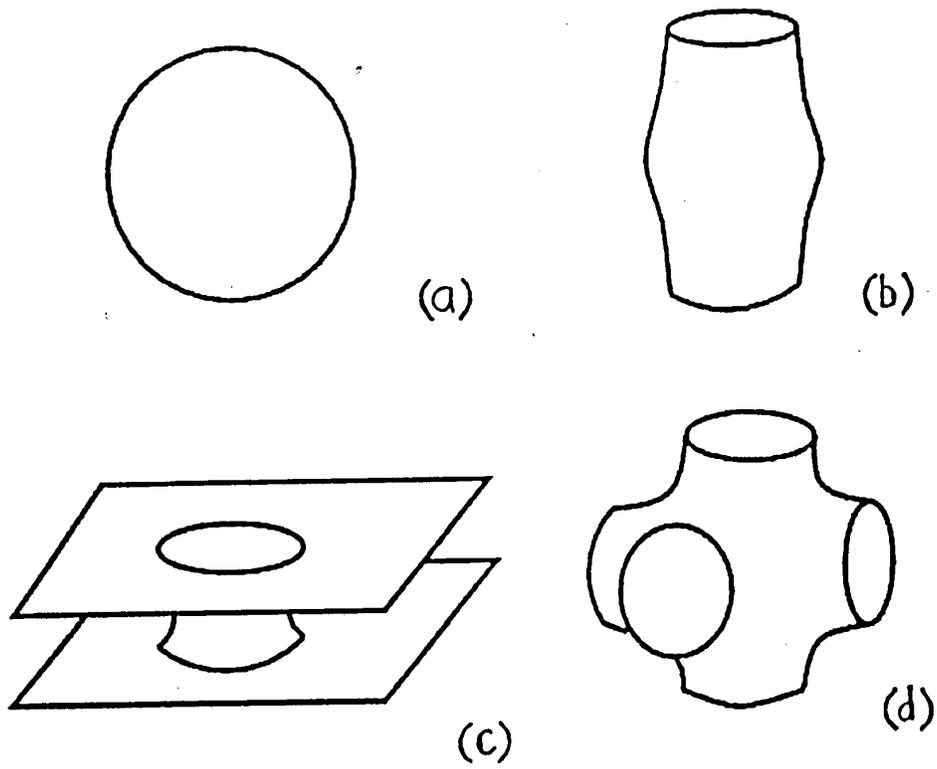,width=14.0cm,height=11cm}
\end{center}
\caption{The Fermi surfaces with Topological Ranks $0$, $1$, $2$
and $3$ respectively.}
\label{drezden8}
\end{figure}

 The pictures on Fig. \ref{drezden8}, a-d represent the pieces of
the Fermi surfaces in ${\mathbb R}^{3}$ with the Topological Ranks
$0$, $1$, $2$ and $3$ respectively.

 As can be seen the genuses of the surfaces
represented on the Fig. \ref{drezden8}, a-d are also equal to
$0$, $1$, $2$ and
$3$ respectively. However, the genus and the Topological Rank are not
necessary equal to each other in the general situation.

 Let us discuss briefly the connection between the genus and the
Topological Rank since this will play the crucial role in further
consideration. 

\vspace{0.5cm}

 It is easy to see that the Topological Rank of the
sphere can be only zero and the Fermi surface consists in this case
of the infinite set of the periodically repeated spheres
${\mathbb S}^{2}$ in ${\mathbb R}^{3}$.

\vspace{0.5cm}

\begin{figure}
\begin{center}
\epsfig{file=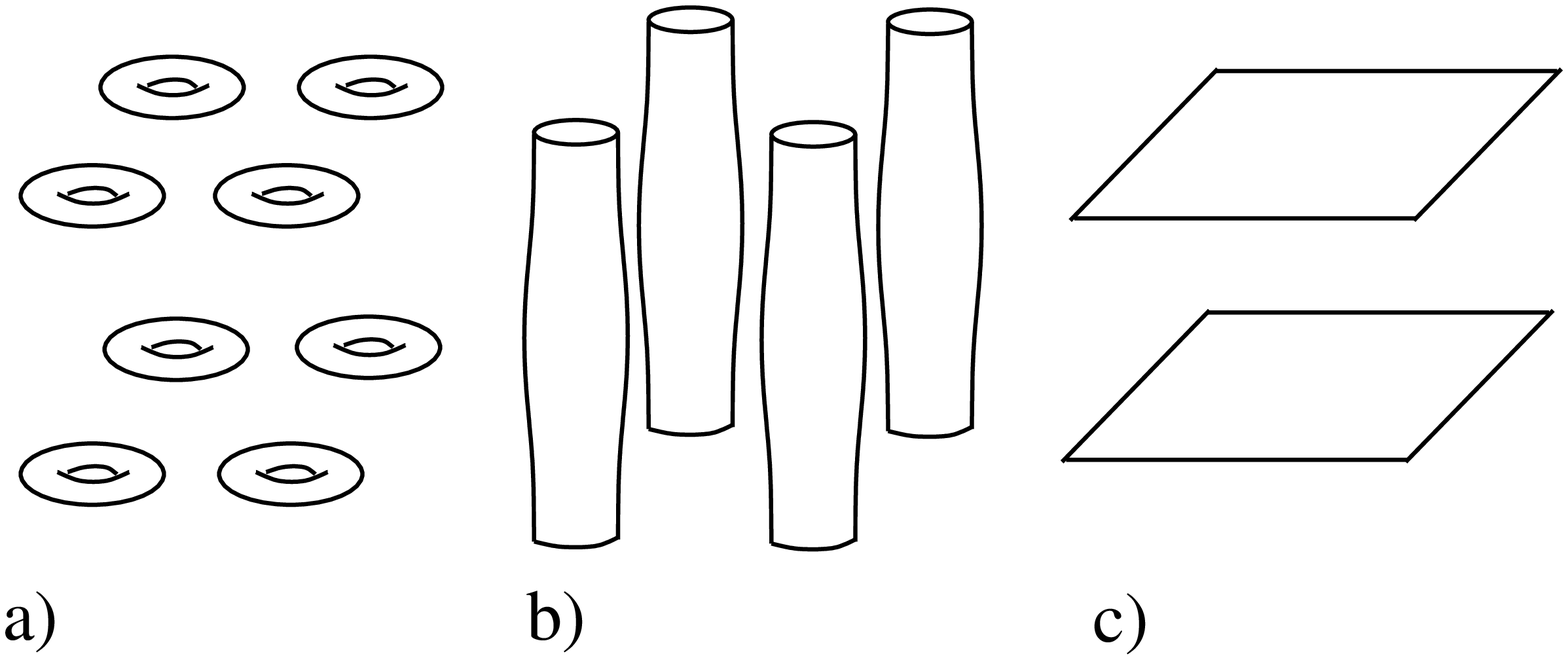,width=14.0cm,height=7cm}
\end{center}
\caption{The periodically repeated tori ${\mathbb T}^{2}$,
periodically repeated  "warped" integral
cylinders and the periodically repeated "warped"
integral planes in ${\mathbb R}^{3}$.
}
\label{drezden9}
\end{figure}

 The Topological Rank of the torus ${\mathbb T}^{2}$ can take
three values $r = 0$, $r = 1$ and $r = 2$.  Indeed, it is easy 
to see that all the three cases of
periodically repeated tori ${\mathbb T}^{2}$ in
${\mathbb R}^{3}$ (Rank $0$), periodically repeated  "warped" 
integral cylinders (Rank $1$) and the periodically repeated "warped"
integral planes (Rank $2$) give the topological 2-dimensional
tori ${\mathbb T}^{2}$ in
${\mathbb T}^{3}$ after the factorization (see Fig. \ref{drezden9}).

 It's not difficult to prove that these are the only possibilities
which we can have for embedding of the 2-dimensional torus 
${\mathbb T}^{2}$ in ${\mathbb T}^{3}$. We just note here that the
mean direction of the "warped periodic cylinder" (embedding of Rank 1) 
can coincide with any reciprocal lattice vector 
$n_{1} {\bf a}_{1} + n_{2} {\bf a}_{2} + n_{3} {\bf a}_{3}$ in
${\mathbb R}^{3}$. Also the "directions" of the corresponding 
"warped planes" (embedding of Rank 2) are always generated by two 
(linearly independent) reciprocal lattice vectors
$m_{1}^{(1)} {\bf a}_{1} + m_{2}^{(1)} {\bf a}_{2} + 
m_{3}^{(1)} {\bf a}_{3}$ and 
$m_{1}^{(2)} {\bf a}_{1} + m_{2}^{(2)} {\bf a}_{2} + 
m_{3}^{(2)} {\bf a}_{3}$. We can see so that both the embeddings
of Rank 1 and Rank 2 of ${\mathbb T}^{2}$ in ${\mathbb T}^{3}$
are characterized by some integer numbers connected with the
reciprocal lattice.

 Let us make also one more remark about the surfaces of Ranks 0,
1 and 2 in this case. Namely the case $r=2$ has actually one
difference from the cases $r=0$ and $r=1$. The matter is
that the plane in ${\mathbb R}^{3}$ is not homological to zero in
${\mathbb T}^{3}$ (i.e. does not restrict any domain of "lower energies") 
after the factorization. We can conclude so that if these planes
appear as the connected components of the physical Fermi surface
(which is always homological to zero) they should always come in pairs, 
$\Pi_{+}$ and $\Pi_{-}$, which are parallel to each other in ${\mathbb 
R}^{3}$. The factorization  of $\Pi_{+}$ and $\Pi_{-}$ gives then the 
two tori ${\mathbb T}^{2}_{+}$, ${\mathbb T}^{2}_{-}$ with the opposite 
homological classes in ${\mathbb T}^{3}$. 

\vspace{0.5cm}

 It can be shown that the Topological Rank of any Fermi surface of
genus $2$ can not exceed $2$ also. The example of the corresponding
immersion of such component with maximal Rank is shown at
Fig. \ref{drezden8}, c and represents the two parallel planes 
connected by cylinders. We will not give the proof of this theorem 
here and just say that this fact plays important role in the 
classification of non-closed electron trajectories on the Fermi
surface of genus 2. Namely, it can be proved that the open
trajectories on the Fermi surface of genus 2 can not be actually
more complicated than the trajectories on the surface of genus 1.
In particular they always have the "topologically regular form"
in the same way  as on the Fermi surface of genus 1 (see later).
Also the same integral characteristics in the cases when this surface
has Rank 1 or 2 as in the case of genus 1 can be introduced for genus
2 (actually for any genus if Rank is equal to 1 or 2).

\vspace{0.5cm}

 At last we say that the Topological Rank of the components with
genus $g \geq 3$ can take any value $r = 0,1,2,3$.

\vspace{0.5cm}

 {\bf Definition 3.} 

{\it We call the open trajectory topologically
regular (corresponding to "topologically integrable" case)
if it lies within the straight line of finite width 
in $\Pi({\bf B})$ and passes through it from $-\infty$ to      
$\infty$. All other open trajectories
we will call chaotic. }

\vspace{0.5cm}

 Let us discuss now the connection between the geometry of the
non-singular electron orbits and the topological properties of the
Fermi surface. We will briefly consider here the simple cases of
Fermi surfaces of Rank 0, 1 and 2 and come then to our basic
case of general Fermi surfaces having the maximal rank
$r = 3$. We have then the following situations:
 
\vspace{0.5cm}

 1) The Fermi surface has Topological Rank $0$.

Easy to see that in this simplest case all the components
of the Fermi surface are compact 
(Fig. \ref{drezden8}a, \ref{drezden9}a) in
${\mathbb R}^{3}$ and there is no open trajectories at all.

\vspace{0.5cm}

  2) The Fermi surface has Topological Rank $1$.

 In this case we
can have both open and compact electron trajectories. However
the open trajectories (if they exist) should be quite simple in  
this case. They can arise only if the magnetic field is orthogonal
to the mean direction of one of the components of Rank $1$
(periodic cylinder) and are periodic with the same integer mean 
direction (Fig. \ref{drezden8}b, \ref{drezden9}b). The corresponding 
sets of the directions ${\bf B}/B$ are just the one-dimensional 
curves and there can not be the open regions on the unit sphere
for which we can find the open trajectories on the Fermi surface.

\vspace{0.5cm}

  3) The Fermi surface has Topological Rank 2.

 It can be easily seen that this case gives much more
possibilities for the existence of open orbits for different
directions of the magnetic field. In particular, this is the
first case where the open orbits can exist for the generic
direction of ${\bf B}$. So, in this case
we can have the whole regions on the unit sphere such that the
open orbits present for any direction of
${\bf B}$ belonging to the
corresponding region. It is easy to see, however, that the open
orbits have also a quite simple description in this case.
Namely, any open orbit (if they exist) lies in this case in
the straight strip of the finite width for any direction of
${\bf B}$ not orthogonal to the integral planes given by the
components of Rank $2$. The boundaries of the corresponding strips
in the planes $\Pi({\bf B})$ (orthogonal to ${\bf B}$) will be
given by the intersection of $\Pi({\bf B})$ with the pairs of
integral planes bounding the corresponding components of Rank $2$.
It can be also shown (\cite{dynn1}, \cite{dynn2}) that every open
orbit passes through the strip from $- \infty$ to  $+ \infty$
and can not turn back. We can see then that all the trajectories
are "topologically regular in this case also.

  According to the remarks above the contribution to the conductivity
given by every family of orbits with the same mean direction
reveals the strong anisotropy when $\omega_{B}\tau \rightarrow \infty$
and coincides in the main order with the formula (\ref{sigmaop})
for the open periodic trajectories.

   Let us say that the trajectories of this type have already
all the features of the general topologically integrable
situation.

\vspace{0.5cm}

 Let us start now with the most general and complicated case
of arbitrary Fermi surface of Topological rank 3.

 We describe first the convenient procedure
(\cite{dynn4},\cite{dynn7}) of reconstruction of the constant
energy surface when the direction of ${\bf B}$ is fixed.  

\begin{figure}
\begin{center}
\epsfig{file=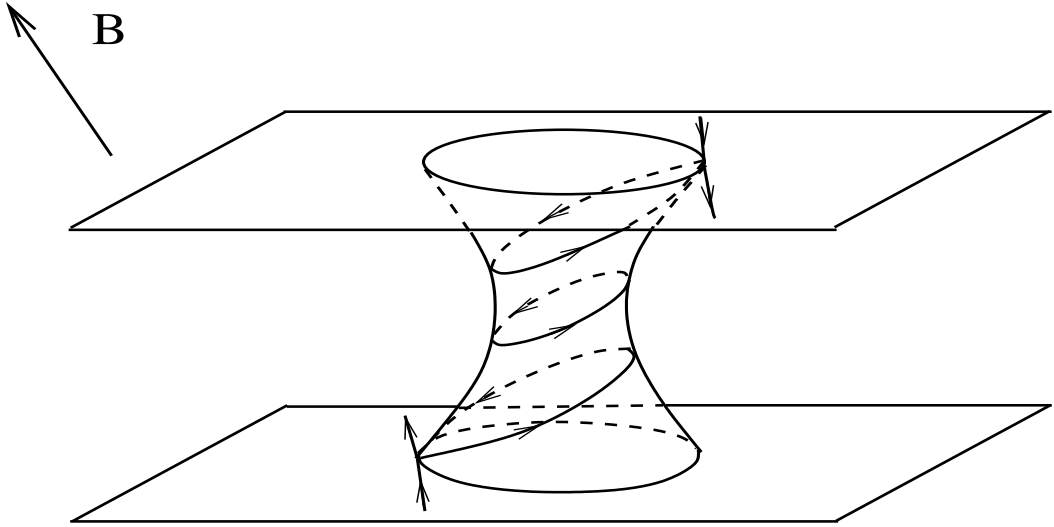,width=14.0cm,height=7cm}
\end{center}
\caption{The cylinder of compact trajectories bounded by the singular
orbits.(The simplest case of just one critical point on the singular
trajectory.)
}
\label{drezden10}
\end{figure}

 We will assume that the system (\ref{dynsyst}) has generically 
only the non-degenerate singularities having
the form of the non-degenerate poles or non-degenerate   
saddle points. The singular trajectories passing through the
critical points (and the critical points themselves) divide 
the set of trajectories into the different parts corresponding
to different types of trajectories on the Fermi surface.
We will not be interested here in the geometry of compact electron 
orbits in the "geometric limit" $\omega_{B} \tau \rightarrow \infty$.
It's not difficult to show that the pieces of the Fermi surface 
carrying the compact orbits can be either infinite or finite 
cylinders in ${\mathbb R}^{3}$ bounded by the
singular trajectories (some of them maybe just points of minimum
or maximum) at the bottom and at the top (see Fig. \ref{drezden10}).

 Let us remove now all the parts containing the non-singular 
compact trajectories from the Fermi surface. The
remaining part

$$S_{F}/({\rm Compact \, Nonsingular \, Trajectories}) \,\,\, = 
\,\,\, \cup_{j} \, S_{j}$$
is a union of the $2$-manifolds $S_{j}$ with boundaries
$\partial S_{j}$ who are the compact singular trajectories.
The generic type in this case is a separatrix orbit with just one
critical point like on the Fig. \ref{drezden10}.

 Easy to see that the open orbit will not be affected at all
by the construction described above and the rest of the Fermi
surface gives the same open orbits as all possible intersections
with different planes orthogonal to ${\bf B}$.

\vspace{0.5cm}

{\bf Definition 4.}

{\it 
 We call every piece $S_{j}$ the {\bf "Carrier of open trajectories"}.
 }

\vspace{0.5cm}
   
\begin{figure}
\begin{center}
\epsfig{file=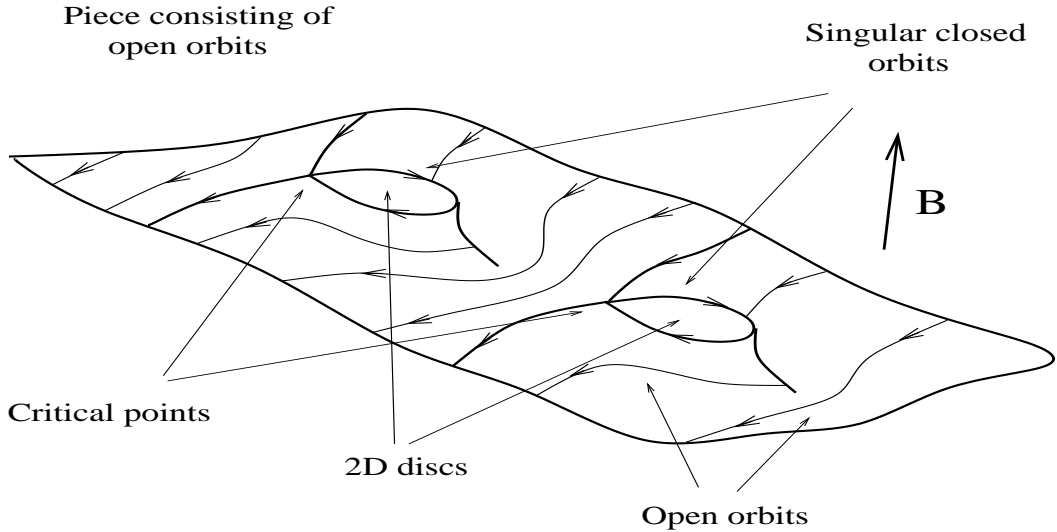,width=14.0cm,height=7cm}
\end{center}
\caption{The reconstructed constant energy surface with removed
compact trajectories and the two-dimensional discs attached to the 
singular trajectories in the generic case of just one critical
point on every singular trajectory.
}
\label{drezden11}
\end{figure}

 Let us fill in the holes by
topological $2D$ discs lying in the planes orthogonal to ${\bf B}$
and get the closed surfaces
 
$${\bar S}_{j} \,\,\, = \,\,\, S_{j} \cup {\rm (2D discs)}$$
(see Fig. \ref{drezden11}).

 This procedure gives again the periodic surface
${\bar S}_{\epsilon}$ after the reconstruction and we can define
the "compactified carriers of open trajectories" both in
${\mathbb R}^{3}$ and ${\mathbb T}^{3}$.

 Easy to see then that the reconstructed surface can be used
instead of the original Fermi surface for the determination of
open trajectories. Let us ask the question: can the reconstructed 
surface be simpler than the original one?

 The answer is positive and moreover it can be proved that 
"generically" the reconstructed surface consists of components
of genus 1 only. This remarkable fact gives the very powerful
instrument for the consideration of open trajectories on the
arbitrary Fermi surface.

  In fact, the proof of the Theorem $1$ was based on
the statement that genus of every compactified carrier of open
orbits ${\bar S}_{j}$ is equal to $1$ in this case.

 Let us formulate now the Theorem of I.A. Dynnikov (\cite{dynn3})
which made the second main breakthrough in the Novikov problem.

\vspace{0.5cm}

{\bf Theorem $2$.} (I.A. Dynnikov, \cite{dynn3}).

{\it Let a generic dispersion relation

$$\epsilon({\bf p}): \,\,\,
{\mathbb T}^{3} \,\,\, \rightarrow {\mathbb R}$$
be given such that for level
$\epsilon({\bf p}) = \epsilon_{0}$ the genus $g$ of some 
carrier of open trajectories ${\bar S}_{i}$ is greater than $1$.
Then there exists an open interval
$(\epsilon_{1}, \epsilon_{2})$ containing $\epsilon_{0}$
such that for all $\epsilon \neq \epsilon_{0}$ in this interval
the genus of carrier of open trajectories is less than $g$. 
}
 
\vspace{0.5cm}

 The Theorem $2$ claims then that only the
"Topologically Integrable case" can be stable with
respect to the small variations of energy level also.

 The formulated theorems permit us to reduce the consideration
of open orbits in any stable situation to the case of the surfaces of 
genus 1 where the Fermi surface can have Topological Rank 0, 1 or 2
only. Easy to see that the Rank 0 can not appear just by definition
in the reconstructed surface ${\bar S}_{\epsilon}$ since it can
contain only the compact trajectories. The Rank 1 is possible in
${\bar S}_{\epsilon}$ only for special directions of ${\bf B}$.
Indeed, the component of Rank 1 has the mean integral direction
in ${\mathbb R}^{3}$ and can contain the open (periodic) trajectories 
only if ${\bf B}$ is orthogonal this integral vector in ${\bf p}$-space.
The corresponding open trajectories is then not absolutely stable
with respect to the small rotations of ${\bf B}$ and can not exist
for the open region on the unit sphere. 

 We can claim then that the only generic situation for 
${\bar S}_{\epsilon}$ is a set of components of Rank 2 which are
the periodic warped planes in this case. The corresponding electron
trajectories can then belong just to "Topologically integrable" case
being the intersections of planes orthogonal to ${\bf B}$ with the
periodically deformed planes in the ${\bf p}$-space.

  The important property of the compactified components of genus 1
arising for the generic directions of ${\bf B}$ is following:
they are all  parallel in average in ${\mathbb R}^{3}$   
and do not intersect each other. This property mentioned in
\cite{novmal1} and called later the  "Topological resonance" plays 
the important role in the physical phenomena connected with
geometry of open trajectories. Such, in particular, all the stable 
topologically regular open trajectories in all planes orthogonal to 
${\bf B}$ have then the same mean direction and give the same 
form (\ref{sigmaop}) of contribution to conductivity in the appropriate 
coordinate system common for all of them. This fact gives the 
experimental possibility to measure the mean direction of non-compact 
topologically regular orbits both in ${\bf x}$ and ${\bf p}$ spaces 
from the anisotropy of conductivity tensor $\sigma^{ik}$.

 Let us say again that the surface ${\bar S}_{\epsilon}$ is the
abstract construction depending on the direction of ${\bf B}$ and
do not exist apriori in the Fermi surface 
$S_{{\epsilon}_{F}}$. The important fact, however, is the 
stability of the surface ${\bar S}_{\epsilon}$ with respect to
the small rotations of ${\bf B}$. This means in particular that the
common direction the components of Rank 2 is locally stable with
respect to the small rotations of ${\bf B}$ which can be then
discovered in the conductivity experiments. From the physical 
point of view, all the regions on the unit sphere  
where the stable open orbits exist can be represented as the 
"stability zones"   
$\Omega_{\alpha}$ such that
each zone corresponds to some integral plane $\Gamma_{\alpha}$
common to all the points of stability zone $\Omega_{\alpha}$.
The plane $\Gamma_{\alpha}$ is then the integral plane in
reciprocal lattice which defines the mean directions of open orbits
in ${\bf p}$-space
for any direction of ${\bf B}$ belonging to $\Omega_{\alpha}$
just as the intersection with the plane orthogonal to ${\bf B}$.
As can be easily seen from the form of (\ref{sigmaop}) this
direction always coincides with the unique direction in
${\mathbb R}^{3}$ corresponding to the decreasing of
conductivity as $\omega_{B} \tau \rightarrow \infty$.

  The corresponding integral planes $\Gamma_{\alpha}$ can then
be given by three integer numbers
$(n^{1}_{\alpha}, n^{2}_{\alpha}, n^{3}_{\alpha})$
(up to the common multiplier) from the equation
$$n^{1}_{\alpha} [{\bf x}]_{1} + n^{2}_{\alpha} [{\bf x}]_{2} +   
n^{3}_{\alpha} [{\bf x}]_{3} = 0$$
where $[{\bf x}]_{i}$ are the coordinates in the basis
$\{{\bf a}_{1}, {\bf a}_{2}, {\bf a}_{3}\}$ of the reciprocal  
lattice, or equivalently

$$n^{1}_{\alpha} ({\bf x}, {\bf l}_{1}) +
n^{2}_{\alpha} ({\bf x}, {\bf l}_{2}) +
n^{3}_{\alpha} ({\bf x}, {\bf l}_{3}) = 0$$
where $\{{\bf l}_{1}, {\bf l}_{2}, {\bf l}_{3}\}$ is the basis
of the initial lattice in the coordinate space.

 We see then that the direction of conductivity decreasing
${\hat \eta} = (\eta_{1}, \eta_{2}, \eta_{3})$ satisfies to relation

$$n^{1}_{\alpha} ({\hat \eta}, {\bf l}_{1}) +
n^{2}_{\alpha} ({\hat \eta}, {\bf l}_{2}) +
n^{3}_{\alpha} ({\hat \eta}, {\bf l}_{3}) = 0$$
for all the points of stability zone $\Omega_{\alpha}$ which
makes possible the experimental observation of numbers
$(n^{1}_{\alpha}, n^{2}_{\alpha}, n^{3}_{\alpha})$.

 The numbers $(n^{1}_{\alpha}, n^{2}_{\alpha}, n^{3}_{\alpha})$
were called in \cite{novmal1} the "Topological Quantum numbers"
of a dispersion relation in metal.

 Let us note, that we can consider now the result of \cite{lifpes1} 
about the "thin spatial net" as the particular case of this general
theorems where the integer planes take the simplest possibility being
the main planes $xy$, $yz$, $xz$. If we introduce now the
"Topological Quantum numbers" for this situation we will have
only the triples $(\pm 1,0,0)$, $(0,\pm 1,0)$ and $(0, 0, \pm 1)$  
for this Fermi surface.

 In general, we can state that the unit sphere should be divided
into the (open) parts where the open orbits are absent at all
on the Fermi level for given directions of ${\bf B}$ and the 
"stability zones" $\Omega_{\alpha}$ where the open orbits exist 
on the Fermi level
and have "topologically regular" form. Every stability zone 
corresponds to the triple of "Topological quantum numbers" giving
the integral direction of periodically deformed two-dimensional
planes in ${\bar S}_{{\epsilon}_{F}}({\bf B})$ which are swept
by the zero eigen-vector of $\sigma^{ik}$ for 
${\bf B} \in \Omega_{\alpha}$. 

\vspace{0.5cm}

 Let us say now that the "Topologically regular" trajectories
are the generic open trajectories but nonetheless they are not the
only possible for rather complicated Fermi surfaces. Namely, for
rather complicated Fermi surfaces and the special directions of
${\bf B}$ the chaotic cases can also arise 
(S.P. Tsarev, I.A. Dynnikov).

  It was first shown by S.P.Tsarev (\cite{tsarev}) that the
more complicated chaotic open orbits can still exist on
rather complicated Fermi surfaces $S_{F}$. Such, the
example of open trajectory which does not lie in any finite
strip of finite width was constructed. However,
the trajectory had in this case the asymptotic direction
even not being restricted by any straight strip of finite
width in the plane orthogonal to ${\bf B}$.
The corresponding asymptotic behavior of conductivity should
reveal also the strong anisotropy properties in the plane   
orthogonal to ${\bf B}$ although the exact form of
$\sigma^{ik}$ will be slightly different from (\ref{sigmaop}) for
this type of trajectories. By the same reason, the asymptotic
direction of orbit can be measured experimentally in this case.

 The more complicated examples of chaotic open orbits were
constructed in \cite{dynn4} for the Fermi surface having genus $3$.
These types of the
open orbits do not have any asymptotic direction in the planes   
orthogonal to ${\bf B}$ and have rather complicated form
"walking everywhere" in these planes. 

 The corresponding contribution to $\sigma^{ik}$ is also very 
different for this kind of trajectories (\cite{maltsev1}). 
In particular, it appears that this contribution becomes zero
in all the directions including direction of ${\bf B}$
for $B \rightarrow \infty$. The total conductivity tensor 
$\sigma^{ik}$ has then only the contribution of compact electron
trajectories in the conductivity along ${\bf B}$ which does
not disappear when $B \rightarrow \infty$. The corresponding 
effect can be observed experimentally as the local minima of the
longitudinal (i.e. parallel to ${\bf B}$) conductivity for the
points of the unit sphere where this kind of trajectories can 
appear. The more detailed description of $\sigma^{ik}$ in this
case can be found in \cite{maltsev1}.

 Let us add that it was proved recently by I.A. Dynnikov that the
measure of chaotic cases on the unit sphere is zero for generic
Fermi surface (\cite{dynn4,dynn7}). 
The systematic investigation of the open orbits was completed
in general after the works \cite{zorich1,dynn3,novmal1,dynn4}  
in \cite{dynn7}. In particular the total picture of
different types of the open orbits for generic dispersion relations
was presented. Let us just formulate here the
main results of \cite{dynn7} in the form of Theorem.

\vspace{0.5cm}

{\bf Theorem $3$} (I.A. Dynnikov, \cite{dynn7}).

{\it Let us fix the dispersion relation
$\epsilon = \epsilon({\bf p})$
and the direction of ${\bf B}$ of irrationality $3$ and consider
all the energy levels for
$\epsilon_{min} \leq \epsilon \leq \epsilon_{max}$. Then:

 1) The open electron trajectories exist for all the energy values
$\epsilon$ belonging to the closed connected energy interval
$\epsilon_{1}({\bf B}) \leq \epsilon \leq \epsilon_{2}({\bf B})$
which can degenerate to just one energy level
$\epsilon_{1}({\bf B}) = \epsilon_{2}({\bf B}) =
\epsilon_{0}({\bf B})$.

 2) For the case of the nontrivial energy interval the set of 
compactified carriers of open trajectories
${\bar S}_{\epsilon}$ is always a disjoint union of
two-dimensional tori ${\mathbb T}^{2}$ in
${\mathbb T}^{3}$ for all
$\epsilon_{1}({\bf B}) \leq \epsilon \leq \epsilon_{2}({\bf B})$.
All the tori
${\mathbb T}^{2}$ for all the energy levels do not intersect
each other and have the same (up to the sign) indivisible
homology class
$c \in H_{2}({\mathbb T}^{3},{\mathbb Z})$, $c \neq 0$.
The number of tori
${\mathbb T}^{2}$ is even for every fixed energy level
and the corresponding covering ${\bar S}_{\epsilon}$ in
${\mathbb R}^{3}$
is a locally stable family of parallel ("warped") integral planes
$\Pi^{2}_{i} \subset {\mathbb R}^{3}$
with common direction given by $c$.
The form of ${\bar S}_{\epsilon}$ described above is locally
stable with the same homology class
$c \in H_{2}({\mathbb T}^{3})$ under
small rotations of ${\bf B}$.
All the open electron trajectories at all the energy levels
lie in the strips of finite width with the same direction and
pass through them. The mean direction of the trajectories is given
by the intersections of planes $\Pi({\bf B})$ with the integral
family $\Pi^{2}_{i}$ for the corresponding "stability zone" on
the unit sphere.

 3) The functions $\epsilon_{1}({\bf B})$, $\epsilon_{2}({\bf B})$
defined for the directions of ${\bf B}$ of irrationality $3$ can
be continuated on the unit sphere $S^{2}$ as the piecewise smooth
functions such that
$\epsilon_{1}({\bf B}) \geq \epsilon_{2}({\bf B})$ everywhere
on the unit sphere.

 4) For the case of trivial energy interval
$\epsilon_{1} = \epsilon_{2} = \epsilon_{0}$ the corresponding
open trajectories may be chaotic. Carrier of the chaotic open
trajectory is homologous to zero in
$H_{2}({\mathbb T}^{3},{\mathbb Z})$ and has genus
$\geq 3$. For the generic energy level $\epsilon = \epsilon_{0}$
the corresponding directions of magnetic fields
belong to the countable union of the codimension $1$ subsets.
Therefore a measure of this set is equal to zero on $S^{2}$.
}

\vspace{0.5cm}

 Let us say that we give here the results connected with generic
directions of ${\bf B}$ and do not consider the special cases when
${\bf B}$ is purely or "partly" rational. The corresponding effects 
are actually simpler then formulated above and can be easily added
to this general picture. Let us give here the references to the survey
articles \cite{novmal2,dynn7,malnov3, malnov4} where all the details
(both from mathematical and physical point of view) can be found.

\section{Quasiperiodic modulations of 2D electron gas and the
general Novikov problem.}

 In this chapter we will give the general impression about the 
quasiperiodic modulations of 2D electron gas and describe the
main topological aspects for the special class of such structures.
Let us say first some words about different modern modulation
techniques and the quasiclassical electron behavior in such
systems. 

 We first point here the holographic illumination of 
high-mobility 2D electron structures ($AlGaAS - GaAs$ 
hetero-junctions) at the temperatures $T \leq 4.2 K$ (see, 
for example \cite{WKPW}). In these experiments the expanded laser
beam was splitten into two parts which gave an interference
picture with the period $a$ on the 2D sample. The illumination 
caused the additional ionization of atoms near the 2D junction which
remains for a rather long time after the illumination. During this
relaxation time the additional periodic potential 
$V({\bf r}) = V(x)$, $V(x) = V(x+a)$ arised in the plane and the 
electron behavior was determined by the orthogonal magnetic field
${\bf B}$ and the potential $V(x)$. 

\begin{figure}
\begin{center}
\epsfig{file=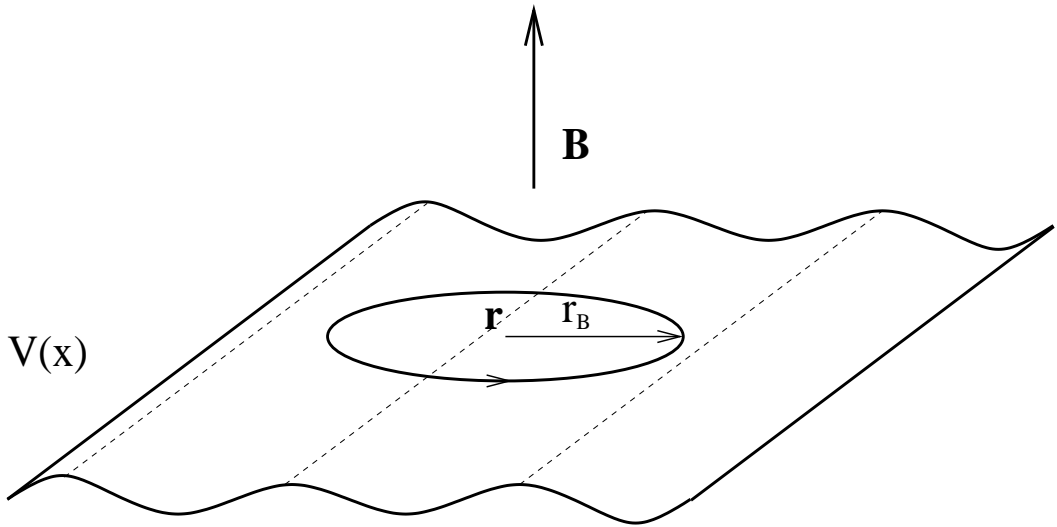,width=14.0cm,height=7cm}
\end{center}
\caption{The averaging of the the potential $V(x)$ over the 
cyclotron orbit with radius $r_{B}$ centered at the point ${\bf r}$.
}
\label{drezden12}
\end{figure}

 The quasiclassical consideration for the case
$|V(x)| \ll \epsilon_{F}$ was first considered by C.W.J. Beenakker
(\cite{beenakker}) for the explanation of "commensurability
oscillations" in such structures found in \cite{WKPW}. According
to this approach the quasiclassical electrons near Fermi level
move around the cyclotron orbits in the magnetic field and drift due
to potential $V(x)$ in the plane. Since only the electrons near
Fermi level $\epsilon_{F}$ play the main role in the conductivity
we can introduce the characteristic cyclotron radius 
$r_{B} = m^{*}v_{F}/eB$ for the Fermi velocity $v_{F}$. The 
corresponding drift of the electron orbits near Fermi will then
be determined by the averaged effective potential $V^{eff}_{B}(x)$
given by the averaging of $V({\bf r}) = V(x)$ over the cyclotron orbit
 with radius $r_{B}$ centered at the point ${\bf r}$ 
(Fig \ref{drezden12}).
 
 The potential $V^{eff}_{B}(x)$ is different from $V(x)$ but has the 
same symmetry and also depends only on $x$. The drift of the cyclotron 
orbits is going along the level curves of $V^{eff}_{B}(x)$ which are
very simple in this case (just the straight lines along the $y$-axis)
and the corresponding velocity $v_{drift}$ is proportional to the 
absolute value of gradient $|V^{eff}_{B}(x)|$ at each level curve.
The analytic dependence of $|V^{eff}_{B}(x)|$ on the value of $B$
(based on the commensurability of $2r_{B}$ with the 
${\rm (integer \, number)} \times a$) was used in \cite{beenakker}
for the explanation of the oscillations of conductivity along
the fringes with the value of $B$. 

 In the paper \cite{GrLonDav} the situation with the double-modulated
potentials made by the superposition of two interference
pictures was also considered. The 
corresponding potential $V({\bf r})$ is double-periodic in 
${\mathbb R}^{2}$ in this case and the same is true for potentials
$V^{eff}_{B}({\bf r})$. The consideration used the same quasiclassical
approach for the potential $V^{eff}_{B}({\bf r})$ based on the
analysis of it's level curves. It was shown then in \cite{GrLonDav} 
that the second modulation should suppress the commensurability 
oscillations in this case which disappear at all for the equal 
intensities of two (orthogonal) interference pictures. 

 Easy to see also that all the open drift trajectories can be only 
periodic in the case of periodic $V^{eff}_{B}({\bf r})$.

\vspace{0.5cm}

 It seems that the situation with the quasiperiodic modulations of 2D
electron gas did not appear in experiments. However, we think that
this situation is also very natural for the technique described above
and can be considered from the point of view of general Novikov
problem. The corresponding approach was developed in \cite{maltsev2}
for the special cases of superpositions of several (3 and 4) 
interference pictures on the plane. Nonetheless, as we already 
mentioned, Novikov problem arise actually also for any picture given
by superposition of several periodic pictures in the plane. The 
corresponding potentials can have many quasiperiods in this case
and the Novikov problem can reveal then much more complicated 
properties (chaotic) than described in \cite{maltsev2}. 

 We are going however to describe here just the main points of 
"topologically regular" behavior in the case of the 
superpositions of 3 and 4 interference pictures which give the
quasiperiodic potentials $V({\bf r})$ and $V^{eff}_{B}({\bf r})$
with 3 and 4 quasiperiods on the plane. Unlike the previous papers
we don't pay here much attention on the analytic dependence on
$B$ and investigate in main the geometric properties of conductivity
in this case. 
 
  Before we start the geometric consideration we want to say
also that the holographic illumination is not the unique way
to produce the superlattice potentials for the two-dimensional
electron gas. Let us
mention here the works \cite{AlBetHen,IASLNT,ISMS,FanStil,TIBAS,
PMMLK,WKPW2,GerWeiWul,DavLar,LarDavLonCus,DavPetLon} where
the different techniques using the biasing of the specially made
metallic gates and the piezoelectric effect were considered. Both
1D and 2D modulated potentials as well as more general
periodic potentials with square and hexagonal geometry
appeared in this situation. 
Actually these techniques give much more
possibilities to produce the potentials of different types   
with the quasiperiodic properties.

\vspace{0.5cm}

\begin{figure}
\begin{center}
\epsfig{file=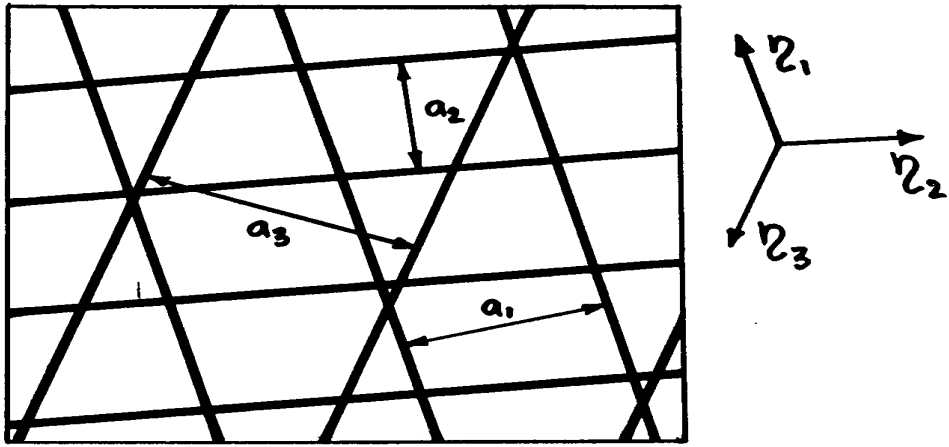,width=14.0cm,height=7cm}
\end{center}
\caption{The schematic sketch of the three  
independent interference pictures on the plane with different
periods and intensities.
}
\label{drezden13}
\end{figure}

 Let us have now three independent interference pictures   
on the plane with three different generic directions
of fringes ${\boldmath \eta}_{1}, 
{\boldmath \eta}_{2}, {\boldmath \eta}_{3}$
and periods $a_{1}, a_{2}, a_{3}$ (see Fig. \ref{drezden13}).

 The total intensity $I({\bf r})$ will be the sum of
intensities

$$I({\bf r}) = I_{1}({\bf r}) + I_{2}({\bf r}) +
I_{3}({\bf r})$$
of the independent interference pictures.

 We assume that there are at least two non-coinciding
directions (say ${\boldmath \eta}_{1}, {\boldmath \eta}_{2}$) 
among the set
$({\boldmath \eta}_{1}, {\boldmath \eta}_{2}, 
{\boldmath \eta}_{3})$.

 It can be shown that the potentials $V({\bf r})$ and
$V^{eff}_{B}({\bf r})$ can be represented in this situation
as the quasiperiodic functions with 3 quasiperiods in the plane.

  Let us introduce now the important definition of the
"quasiperiodic group" acting on the potentials described
above.

\vspace{0.5cm} 

 {\bf Definition 5.} 

{\it Let us fix the directions 
${\boldmath \eta}_{1}, {\boldmath \eta}_{2}, 
{\boldmath \eta}_{3}$ 
and periods 
$a_{1}, a_{2}, a_{3}$ of the interference fringes on picture
\ref{drezden13} and consider all independent parallel shifts of
positions of different interference pictures in ${\mathbb R}^{2}$.
We will say that all the potentials $V^{\prime}({\bf r})$
(and the corresponding $V^{eff \, \prime}_{B}({\bf r})$) made
in this way are related by the transformations of a quasiperiodic
group.}

\vspace{0.5cm}

 According to the definition the quasiperiodic group is a 
three-parametric Abelian group isomorphic to the 3-dimensional 
torus ${\mathbb T}^{3}$ due to the periodicity of every interference
picture.\footnote{Easy to see that the quasiperiodic group contains
the ordinary translations as the algebraic subgroup.}

\vspace{0.5cm}

 We will say that potential $V({\bf r})$ is generic if it has no
periods in ${\mathbb R}^{2}$. We say that potential $V({\bf r})$
is periodic if it has two linearly independent periods in
${\mathbb R}^{2}$ and that $V({\bf r})$ is "partly periodic"
if it has just one (up to the integer multiplier) period in
${\mathbb R}^{2}$.

\vspace{0.5cm}

 It can be also shown that the quasiperiodic group does not change 
the "periodicity" of potentials $V({\bf r})$, $V^{eff}_{B}({\bf r})$.

 Let us say now that the results for Novikov problem can be applied
also in this situation. We will formulate here
the main results for the generic potentials $V({\bf r})$
(the special additional features can be found in \cite{maltsev2}).
Let us formulate here the theorem from \cite{maltsev2} about the
drift trajectories for the generic potentials of this kind based
on the topological theorems for Novikov problem in 3-dimensional 
case (formulated above).

\vspace{0.5cm}

{\bf Theorem 4.} (\cite{maltsev2})

{\it Let us fix the value of $B$ and consider
the generic quasiperiodic potential $V_{B}^{eff}({\bf r})$ 
made by three interference pictures and
taking the values in some interval
$\epsilon_{min}(B) \leq V_{B}^{eff}({\bf r})
\leq \epsilon_{max}(B)$.
Then:

 1) Open quasiclassical trajectories $V_{B}^{eff}({\bf r}) = c$
always exist either in the connected energy interval

$$\epsilon_{1}(B) \leq c \leq \epsilon_{2}(B)$$
$(\epsilon_{min}(B) < \epsilon_{1}(B) <
\epsilon_{2}(B) < \epsilon_{max}(B))$
or just at one energy value $c = \epsilon_{0}(B)$.

 2) For the case of the finite interval
($\epsilon_{1}(B) < \epsilon_{2}(B)$) all the non-singular open
trajectories correspond to topologically regular case, i.e.
lie in the straight strips of the finite width and pass 
through them. All the strips have the same mean directions
for all the energy levels
$c \in [\epsilon_{1}(B), \epsilon_{2}(B)]$
such that all the open trajectories are in average parallel to
each other for all values of $c$.

 3) The values $\epsilon_{1}(B)$, $\epsilon_{2}(B)$ or
$\epsilon_{0}(B)$ are the same for all the generic potentials 
connected by the "quasiperiodic group".

 4) For the case of the finite energy interval
($\epsilon_{1}(B) < \epsilon_{2}(B)$) all the non-singular open
trajectories also have the same mean direction for all the
generic potentials connected by the "quasiperiodic group"
transformations. }

\vspace{0.5cm}

 We see again that the "topologically regular" open trajectories
are also generic for this situation as previously.

 Let us consider now the asymptotic behavior of conductivity   
tensor when $\tau \rightarrow \infty$ (mean free electron motion time).
We will consider here only the "topologically regular" case. Let us 
point out that the full conductivity tensor can be represented as
the sum of two terms

$$\sigma^{ik}_{0}(B) = \sigma^{ik}_{0}(B) +
\Delta \sigma^{ik}(B) $$

 In the approximation of the drifting cyclotron orbits
the parts $\sigma^{ik}_{0}(B)$ and $\Delta \sigma^{ik}(B)$
can be interpreted as caused respectively by the
(infinitesimally small) difference in the electron
distribution function on the same cyclotron orbit (weak 
angular dependence) and the (infinitesimally small) difference  
in the occupation of different trajectories by the centers of
cyclotron orbits at different points of  
${\mathbb R}^{2}$ (on the same energy level) as the linear
response to the (infinitesimally) small external field ${\bf E}$.
   
 The first part $\sigma^{ik}_{0}(B)$ has the standard asymptotic
form:

$$\sigma^{ik}_{0}(B) \sim {n e^{2} \tau \over m^{eff}}
\left( \begin{array}{cc}
(\omega_{B} \tau)^{-2} & (\omega_{B} \tau)^{-1} \cr
(\omega_{B} \tau)^{-1} & (\omega_{B} \tau)^{-2}
\end{array} \right) $$  
for $\omega_{B} \tau \gg 1$ due to the weak angular dependence
$( \sim 1/\omega_{B} \tau )$ of the distribution function on the
same cyclotron orbit. We have then that the corresponding 
longitudinal conductivity decreases for
$\tau \rightarrow \infty$ in all the directions in
${\mathbb R}^{2}$ and the corresponding condition is    
just $\omega_{B} \tau \gg 1$ in this case.

 For the part $\Delta \sigma^{ik}(B)$ the limit
$\tau \rightarrow \infty$ should, however, be considered  
as the condition that every trajectory is passed for rather
long time by the drifting cyclotron orbits to reveal its global
geometry. Thus another parameter $\tau/\tau_{0}$ where
$\tau_{0}$ is the characteristic time of completion of close   
trajectories should be used in this case and we should put
the condition $\tau/\tau_{0} \gg 1$ to have the asymptotic
regime for $\Delta \sigma^{ik}(B)$.
In this situation the difference between the
open and closed trajectories  plays the main role and the
asymptotic behavior of conductivity can be calculated in
the form analogous to that used in \cite{lifazkag,lifpes1,lifpes2}
for the case of normal metals. Namely:

$$ \Delta \sigma^{ik}(B) \sim
{n e^{2} \tau \over m^{eff}}
\left( \begin{array}{cc}
({\tau_{0}/\tau})^{2} & \tau_{0}/\tau \cr
\tau_{0}/\tau & ({\tau_{0}/\tau})^{2}
\end{array} \right) $$
in the case of closed trajectories and

$$ \Delta \sigma^{ik}(B) \sim
{n e^{2} \tau \over m^{eff}}
\left( \begin{array}{cc}
* & \tau_{0}/\tau \cr
\tau_{0}/\tau & ({\tau_{0}/\tau})^{2}
\end{array} \right) $$
($* \sim 1$) for the case of open topologically regular
trajectories if the $x$-axis coincides with the mean direction 
of trajectories.

  The condition $\tau/\tau_{0} \gg 1$ is much stronger
then $\omega_{B} \tau \gg 1$ in the situation described above
just according to the definition of the slow drift of the cyclotron
orbits. We can keep then just this condition in our further
considerations and assume that the main part of conductivity is
given by $\Delta \sigma^{ik}(B)$ in this limit. Easy to see
also that the magnetic field $B$ should not be "very strong"
in this case.

  According to the remarks above we can write now the main part
of the conductivity tensor $\sigma^{ik}(B)$ in the limit
$\tau \rightarrow \infty$ for the case of topologically regular
open orbits. Let us take the $x$ axis along the mean direction
of open orbits and take the $y$-axis orthogonal to $x$. The  
asymptotic form of $\sigma^{ik}$, $i, k = 1, 2$ can then
be written as:

\begin{equation}
\label{sigmaik}
\sigma^{ik} \sim
{n e^{2} \tau \over m^{eff}}
\left( \begin{array}{cc}
* & \tau_{0}/\tau \cr
\tau_{0}/\tau & ({\tau_{0}/\tau})^{2}
\end{array} \right) \,\,\, , \,\,\,
\tau_{0}/\tau \rightarrow 0
\end{equation}
where $*$ is some value of order of 1 (constant as
$\tau_{0}/\tau \rightarrow 0$).

 The asymptotic form of $\sigma^{ik}$ makes possible the   
experimental observation of the mean direction of topologically
regular open trajectories if the value $\tau/\tau_{0}$ is rather
big.

\vspace{0.5cm}

 Let us introduce now the "topological numbers" characterizing
the regular open trajectories analogous to introduced in 
\cite{novmal1} for the case of normal metals. We will give first
the topological definition of these numbers using the action of the
"quasiperiodic group" on the quasiperiodic potentials 
(\cite{maltsev2}).

 We assume that we have the
"topologically integrable" situation where the topologically
regular open trajectories exist in some finite energy interval
$\epsilon_{1}(B) \leq c \leq \epsilon_{2}(B)$. According to
Theorem 4 the values $\epsilon_{1}(B)$, $\epsilon_{2}(B)$ and
the mean directions  
of open trajectories are the same for all the potentials
constructed from our potential with the aid of the
"quasiperiodic group". It follows also from the topological
picture that all the topologically regular trajectories are
absolutely stable under the action of the "quasiperiodic group"
for the generic $V^{eff \, \prime}_{B}({\bf r})$ and can just 
"crawl" in the plane for the continuous action of
such transformations.

 We take the first interference picture
($({\boldmath \eta}_{1}, a_{1})$) and shift
continuously the interference fringes in the direction
of $grad \, X({\bf r})$ (orthogonal
to ${\boldmath \eta}_{1}$) to the distance $a_{1}$ 
keeping two other pictures unchanged. Easy to see that we will have
at the end the same potentials $V(x,y)$ and $V_{B}^{eff}(x,y)$ 
due to the periodicity
of the first interference picture with period $a_{1}$.
Let us fix now some energy level
$c \in (\epsilon_{1}(B), \epsilon_{2}(B))$ and look at the evolution
of non-singular open trajectories (for $V_{B}^{eff}(x,y)$) 
while making our transformation.
We know that we should have the parallel open trajectories in
the plane at every time and the initial picture should coincide
with the final according to the construction. The form of
trajectories can change during the process but their mean  
direction will be the same according to Theorem 4 
("Topological resonance").

\begin{figure}
\begin{center}
\epsfig{file=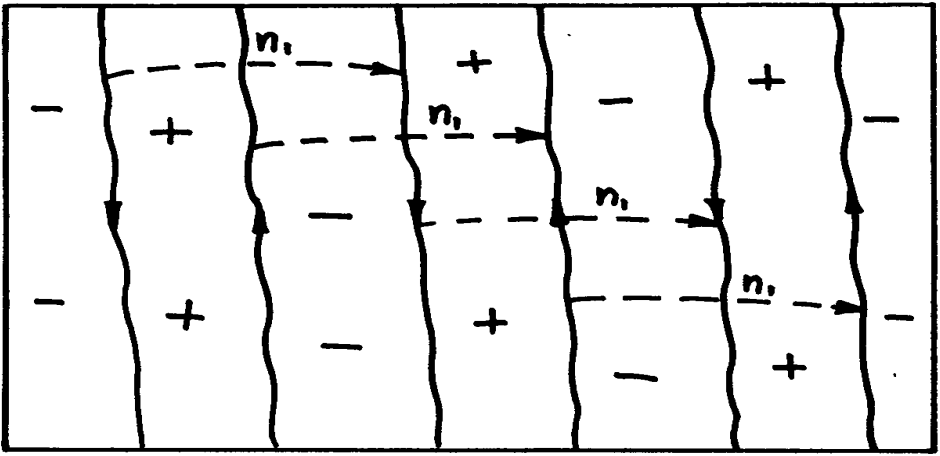,width=14.0cm,height=7cm}
\end{center}
\caption{The shift of "topologically regular"
trajectories by a continuous transformation generated by the
special path in the "quasiperiodic group".
}
\label{drezden14}
\end{figure}

 We can claim then that every open trajectory will be "shifted"
to another open trajectory of the same picture by our continuous
transformation. It's not difficult to prove that all the
trajectories will then be shifted by the same number of positions
$n_{1}$ (positive or negative) which depends on the potential
$V_{B}^{eff}(x,y)$ (Fig. \ref{drezden14}).

 The number $n_{1}$ is always even since all the trajectories  
appear by pairs with the opposite drift directions.

 Let us now do the same with the second and the third sets of
the interference fringes and get an integer triple
$(n_{1}, n_{2}, n_{3})$ which is a topological characteristic
of potential $V_{B}^{eff}(x,y)$ (the "positive" direction of
the numeration of trajectories should be the same for all these
transformations).
 
 The triple $(n_{1}, n_{2}, n_{3})$ (defined up to the common  
sign) can be represented as:

$$(n_{1}, n_{2}, n_{3}) \, = \, M \, (m_{1}, m_{2}, m_{3}) $$
where $M \in {\mathbb Z}$ and $(m_{1}, m_{2}, m_{3})$
is the indivisible integer triple.

\begin{figure}
\begin{center}
\epsfig{file=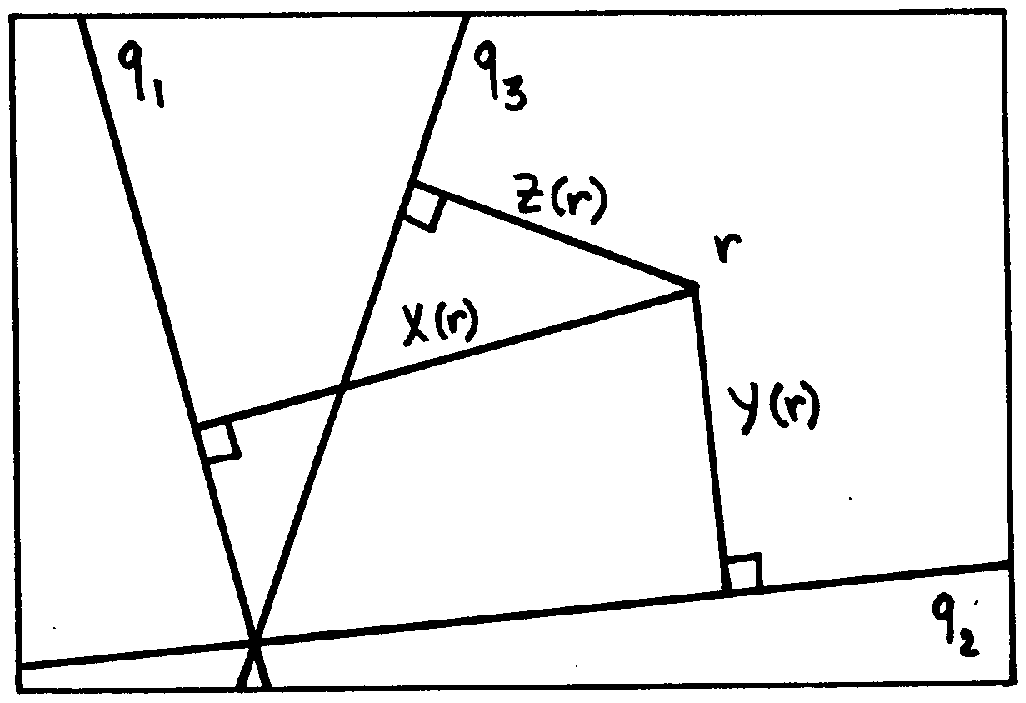,width=14.0cm,height=7cm}
\end{center}
\caption{The function $X({\bf r})$, $Y({\bf r})$
and $Z({\bf r})$ on the plane.
}
\label{drezden15}
\end{figure}

 The numbers $(m_{1}, m_{2}, m_{3})$ play now the role of 
"Topological numbers" for this situation. Let us say that for direct
experimental observation of these numbers the connection between these 
numbers and the mean direction of the "Topologically regular" 
trajectories can play important role. Let us describe here this
connection:

 Let us draw three straight lines $q_{1}$, $q_{2}$, $q_{3}$
with the directions 
${\boldmath \eta}_{1}, {\boldmath \eta}_{2}, {\boldmath \eta}_{3}$
(Fig. \ref{drezden13})
and choose the "positive" and "negative" half-planes for every line
$q_{i}$ on the plane. Let us consider now three linear functions
$X({\bf r})$, $Y({\bf r})$, $Z({\bf r})$ on the plane which are
the distances from the point ${\bf r}$ to the lines
$q_{1}$, $q_{2}$, $q_{3}$ with the signs $"+"$ or $"-"$
depending on the half-plane for the corresponding line $q_{i}$
(Fig. \ref{drezden15}). Let us choose here the signs $"+"$ or $"-"$
such that the gradients of $X({\bf r})$, $Y({\bf r})$, $Z({\bf r})$
coincide with directions of shifts of the corresponding interference
pictures in the definition of $(m_{1}, m_{2}, m_{3})$.

\vspace{0.5cm}

{\bf Theorem 6} (\cite{maltsev2})

{\it Consider the functions

$$X^{\prime} ({\bf r}) = X({\bf r})/a_{1} \,\,\, , \,\,\,    
Y^{\prime} ({\bf r}) = Y({\bf r})/a_{2} \,\,\, , \,\,\,     
Z^{\prime} ({\bf r}) = Z({\bf r})/a_{3} $$
in ${\mathbb R}^{2}$. The mean direction of the regular open
trajectories is given by the linear equation:

\begin{equation}
\label{dircond}
m_{1} X^{\prime}(x,y) + m_{2} Y^{\prime}(x,y) +
m_{3} Z^{\prime}(x,y) = 0
\end{equation}
where $(m_{1}, m_{2}, m_{3})$ is the indivisible integer triple
introduced above. }

\vspace{0.5cm}

 Let us say now about the situation with 4 independent sets of
interference fringes in the plane (see also \cite{maltsev2}). 
In general we get here the
quasiperiodic potentials $V({\bf r})$, $V^{eff}_{B}({\bf r})$
with 4 quasiperiods. The situation in this case is more complicated 
than in the case $N = 3$ and no general classification of open 
trajectories exists at the time. At the moment just the theorem 
analogous to Zorich result can be formulated in this situation
(S.P. Novikov, \cite{novikov5}). According to Novikov theorem
we can claim just that the "small perturbations" of purely
periodic potentials having 4 quasiperiods have the 
"topologically regular" level curves like in the previous case.

 Let us say that the purely periodic potentials $V({\bf r})$
give the everywhere dense set in the space of parameters
${\boldmath \eta}_{1}$, ${\boldmath \eta}_{2}$, 
${\boldmath \eta}_{3}$, ${\boldmath \eta}_{4}$, 
$a_{1}$, $a_{2}$, $a_{3}$, $a_{4}$ and can be found in any
small open region of this space. Novikov theorem claims then
that every potential of this kind can be surrounded by the
"small open ball" in the space of parameters
${\boldmath \eta}_{1}$, ${\boldmath \eta}_{2}$,
${\boldmath \eta}_{3}$, ${\boldmath \eta}_{4}$,
$a_{1}$, $a_{2}$, $a_{3}$, $a_{4}$ where the open level curves
will always demonstrate the "topologically regular" behavior.
The set of potentials thus obtained has the finite measure among
all potentials and the "topologically regular" open trajectories can 
be found with finite probability also in this case. However, we
don't claim here that the chaotic behavior has measure zero for
4 quasiperiods and moreover we expect the nonzero probability also 
for the chaotic trajectories in this more complicated case.

 The topologically regular cases demonstrate here the same 
"regularity properties" as in the previous case including the
"Topological numbers". Thus, we can introduce in the same way
the action of the quasiperiodic group on the space of potentials 
with 4 quasiperiods and define in the same way the 4-tuples 
$(m_{1}, m_{2}, m_{3}, m_{4})$ of integer numbers characterizing
the topologically regular cases in this situation. 

 Also the analogous theorem about mean directions of the regular 
trajectories can be formulated in this case. Namely, if we introduce
the functions $X({\bf r})$, $Y({\bf r})$, $Z({\bf r})$, $W({\bf r})$
in the same way as for the case of 3 quasiperiods (above) and
the corresponding functions 

$$X^{\prime}({\bf r}) = X({\bf r})/a_{1} \,\,\, , \,\,\, \dots
\,\,\, , \,\,\, W^{\prime}({\bf r}) = W({\bf r})/a_{4} $$
we can write the equation for the mean direction of open trajectories
on the plane in the form:

$$m_{1} X^{\prime}({\bf r}) + m_{2} Y^{\prime}({\bf r}) +
m_{3} Z^{\prime}({\bf r}) + m_{4} W^{\prime}({\bf r}) = 0 $$

 The numbers $(m_{1}, m_{2}, m_{3}, m_{4})$ are stable with respect
to the small variations of 
${\boldmath \eta}_{1}$, ${\boldmath \eta}_{2}$,
${\boldmath \eta}_{3}$, ${\boldmath \eta}_{4}$,
$a_{1}$, $a_{2}$, $a_{3}$, $a_{4}$ (and the intensities of
the interference pictures $I_{1}$, $I_{2}$, $ I_{3}$, $I_{4}$ and 
correspond again to some "stability zone' in this space of
parameters.

\vspace{0.5cm}

 Let us say now some words about the limit of Novikov problem
for the large values of $N$. Namely, the following problem can be
formulated:

 Give a description of global geometry of the open level curves
of quasiperiodic function $V({\bf r})$ in the limit of large numbers 
of quasiperiods.

 We can claim that the open level curves should exist here also in 
the connected energy interval $[\epsilon_{1}, \epsilon_{2}]$ on the
energy scale which can degenerate just to one point $\epsilon_{0}$.
\footnote{The proof given in \cite{dynndiss} for the case of 3
quasiperiod works actually for any $N$.} We expect that the
"topologically regular" open trajectories can exist also in this 
case. However the probability of "chaotic behavior" should increase
for the cases of large $N$ which is closer now to random potential
situation. The corresponding behavior can be considered then as the
"percolation problem" in special model of random potentials given
by quasiperiodic approximations. Certainly, this model can be quite 
different from the others. Nevertheless, we expect the similar
behavior of the chaotic trajectories for rather big $N$ also in this 
rather special model. This area, however, is still under investigation
by now.

\end{document}